# Extending the Width of Short Aperture Data
# by Data-Driven Extrapolation

Gerard T. Schuster* and Jing Li[†]
*Geology and Geophysics Department, U of Utah (gerard.schuster2022@gmail.com)
[†]College of Geo-exploration Science and Technology, Jilin University (inter.lijing@gmail.com)

## ABSTRACT

It is commonly believed that it is infeasible to estimate the phase velocity of a surface wave with a recording aperture less than $\lambda/2$ in length, where $\lambda$ is the horizontal wavelength of the propagating wave. We show both theoretically and computationally that the width of the recording aperture can be extended using the recorded data at longer time intervals. This leads to more reliable estimates of apparent velocities for both body waves and surface waves. Here, we assume single-mode plane waves sweeping across the recording array with a linear moveout.

## INTRODUCTION

Rayleigh waves recorded at lower frequencies can be inverted to provide deeper images of the S-velocity profile and the apparent velocity of diving waves can be used to infer the sub surface velocity model as a function of depth (Aki and Richards, 1980). For surface waves, the typical inversion strategy is to estimate the phase velocity of the arrival at a specified frequency, and then create the observed dispersion curve in phase-velocity and frequency coordinates. S-velocity models are then found to provide the theoretical dispersion curve that best matches the observed one. One of the perceived limitations of these methods is that it is infeasible to estimate the phase velocity of a surface wave with a recording aperture less than $\lambda/2$ in length, where $\lambda$ is the wavelength of the surface wave at a fixed frequency. We now show that there is no such limitation by using the surface wave information over longer time windows.

## EXTENSION OF SHORT APERTURE RECORDINGS

Assume that the recording aperture of the sampled surface wave at a fixed frequency $\omega$ is $\lambda/4$. For the geophone configuration at Vandenberg Space Force base (VSFB) on July 11, 2025, synthetic 1 Hz seismograms are depicted in Figure 1a spatially sampled at 0.018 km. For convenience we assume a single mode of surface wave. In this example the aperture width X of the recording array is 0.18 km and the wavelength is 1 km, so that the aperture width of X =0.18 ≈ 5 is less than 1/5 the wavelength of 1 km. Thus, an automatic phase-estimation method such as semblance (Yilmaz, 2001) will likely fail because with noisy data there is a wide range of slopes that can be fitted to the events in Figure 1b. A typical strategy is to avoid estimating phase velocities from events where the recording aperture is less than $\lambda/2$. Unfortunately this prevents short aperture data from estimating the apparent velocity of low-frequency arrivals.

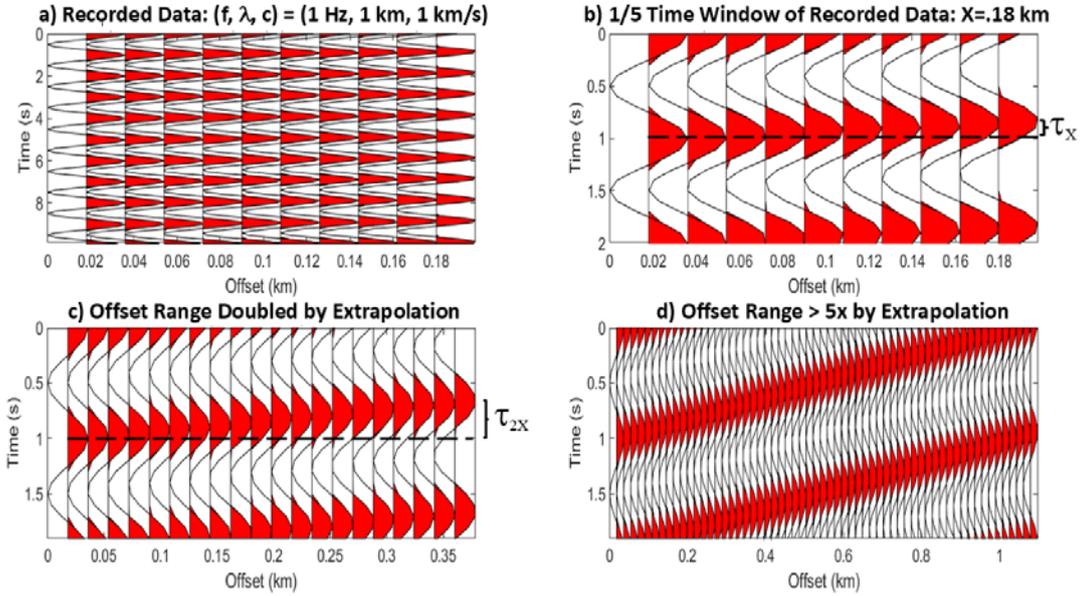

Figure 1: Monofrequency synthetic data: a) traces "recorded" at 10 geophones, b) $0 - 2$ sec time-window image from a), c) windowed data from b) appended with extrapolated traces (see equation 3), and d) data with a recording width of 0.18 km in a) extrapolated to 1 km. The time shift $\tau X$ in b) is used for data extrapolation in equation 3.

As an example, Figure 2b depicts the semblance estimate of apparent velocities com puted from the well-sampled CSG in Figure 2a. In this case the semblance values are closely centered around the correct moveout velocities of c =1.5, 2.0, and 3.0km/s. In comparison, the CSG with 1/4 the aperture width in Figure 2c results in the semblance image in Figure 2d where there is a wide variation of moveout velocities centered around the correct ones. This is not surprising, since a flat linear event with period $k = \text{T0}/2$ at an offset X will have a slope variation between $\pm\frac{\text{T0}}{2\text{X}}$ (The dominant semblance energy is computed when summing the trace amplitudes along slopes between $\pm\frac{\text{T0}}{2\text{X}}$ .), which means that the dominant variation of semblance energy will be inversely proportional to offset X. See Appendix for an example.

To digitally extend the aperture we can use a time window in the seismograms to extrapolate the offset range of recorded seismograms. This can be done by recognizing that a harmonic plane wave, such as a propagating harmonic surface wave with a single mode, is given as

$$f(x,t)_\omega = \cos(kx - \omega t),$$

(1)

where the wavenumber $k = \omega/c$, $\omega$ is angular frequency, and c is the horizontal propagation velocity. Here $f(x,t)_\omega$ is denoted as the window of harmonic data that starts at $t$ and ends at $2T_o$, where $T_o$ is the period. The offset range of this data will be denoted as $1/5\ \lambda$, as depicted in Figure 1b.

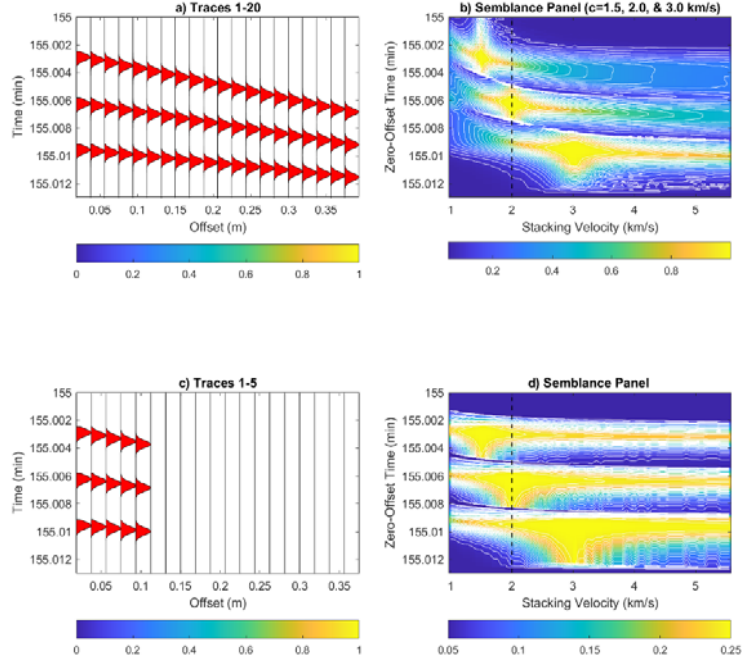

Figure 2: Synthetic data: a) seismograms at 20 geophones, b) semblance image from a), c) seismograms at 5 geophones, and d) semblance image computed from c). The linear moveout velocity of the three vents are c =1.5, 2.0, and 3.0km/s.

To extend the recording aperture, assume that the harmonic data $f(x,t)_\omega$ are recorded for $0 \le x \le X$ and a time slice $f(x,t)_\omega$ is given by

$$
f(x, t - \tau_X)_\omega |_{x=0}^{X} = \overbrace{\cos(kx - \omega(t - \tau_X))}^{time-window\ of\ recorded\ data\ 0 \le x \le X},
$$
(2)

$$
= \cos(kx + \overbrace{\omega\tau_X}^{\omega\tau_X = \omega\tau_X c/c = k\lambda_X = kX} - \omega t),
$$

$$
= \overbrace{\cos(k(x + X) - \omega t)}^{virtual\ data\ recorded\ at\ extended\ aperture\ X \le x \le 2X},
$$
(3)

where $\tau X$ is the time shift between the peak at x = 0 in Figure 1b and its shifted copy at the end of recording array at x = X; this time shift $\tau X$ is used to compute the corresponding apparent wavelength $\lambda X$

= cτX = X that is equal to the aperture width of X. Here, cos(k(x+X)−ωt) is interpreted as virtual data that is virtually recorded at the offset range between x = X to x =2X, which virtually doubles the aperture width as demonstrated in extrapolating Figure 1b to Figure 1c; we will call this virtual data the extrapolated data. The data described in equation 2 is a time-windowed portion of the recorded harmonic data, which can be concatenated to the right of Figure 1b to get Figure 1c at double the aperture.

This extrapolation procedure can be repeated in equation 3 by measuring the time shift τ2X in Figure 1c and using equation 3 to further extend the aperture. Figure 1d shows an example where the aperture width is more than 5 times that Figure 1b. The significance of the extrapolated data is that it is now possible to get more reliable values of phase velocity for this frequency using a semblance-like method if a reliable estimate of τ is estimated from the data.

## WORFKOW

The workflow for the extrapolation procedure is the following, where the seismograms d(x,t) have been recorded from x =0 to x = X.

1. Apply an FFT to a trace d(x,t) at offset x to get $\tilde{d}(x,\omega)$. For convenience, symmetrize d(x,t) such that d(x,t)=d(x,−t) so that its spectrum $\tilde{d}(x,\omega)$ is real. In practice, it might be better to use a narrow-band Fourier transform or a short-time Fourier transform to obtain $\tilde{d}(x,\omega)$, where ω is the center frequency of the narrow-band filter.

2. Define the extrapolated data between the offsets X to 2X as

$$
\begin{aligned}
\tilde{e}(x+X,t)_\omega|_{x=0}^X &= \tilde{d}(x,\omega)\cos(kx-\omega(t-\tau_X)), \\
&= \tilde{d}(x,\omega)\cos(k(x+X)-\omega t), \quad (4)
\end{aligned}
$$

where $\tilde{e}(x+X,t)_\omega|_{x=0}^X$ is the virtual data in the extended aperture that fills in the recording gap at X<x2X. Ideally, the amplitude $\tilde{d}(x,\omega)$ in the above equation should be the same as the amplitude $\tilde{d}(x+X,\omega)$ ˜ but since we are interested in semblance values the approximation $\tilde{d}(x+X,\omega) \approx \tilde{d}(x,\omega)$ should not affect normalized traces.

3. For surface waves with a single mode(At low frequencies, the surface wave is dominated by the fundamental mode), then a semblance procedure can be applied to $\tilde{e}(x+X,t)_\omega|_{x=0}^X$ concatenated with the harmonic data recorded from $0 \leq x \leq X$ to use semblance to compute the phase velocity at the frequency ω.

In theory, a reliable estimation of the phase velocity as a function of low frequencies can be obtained as long as a reliable estimate of τ is obtained from the short offset data.

## Extrapolating Diving Wave Events Generated by a Falcon 9 Launch

A Falcon 9 launch on July 11, 2024 from VSFB was seismically recorded by vertical component geophones located about 7 km from the launch pad (Schuster et al., 2025). Figure 3a depicts the 9 seismograms with coherent events interpreted as diving waves that penetrated deeper than 1 km into the crust. Evidence for this interpretation included the estimate of the apparent velocity in Figure 3b obtained by a semblance method (Yilmaz, 2001). The apparent velocity is estimated to be around 2.1 km/s, but the wide contours suggest a wide variation of possible velocities. This large variance of velocities is characteristic of the short offset of the recording array, which is about a 1/3 of the apparent wavelength of these diving waves.

To reduce this large variance, we use the extrapolation method described in the previous section. Equivalently, diving events in Figure 3a can be windowed between 155.285-155.287 minutes, and a time-shifted version can be appended to traces 1-9 to give the 18 traces in Figure 4c. The time shift τX between the near- and far-offset traces for the windowed events is measured from Figure 4a.

The semblance images associated with Figure 4a and Figure 4c are shown to their right. It is obvious that the wide contours, or variation of stacking velocities, in Figure 4d are significantly thinner than those in Figure 4b. This, of course, results from doubling the aperture of recorded traces where the variation of reasonable stacking velocities is inversely proportional to aperture width.

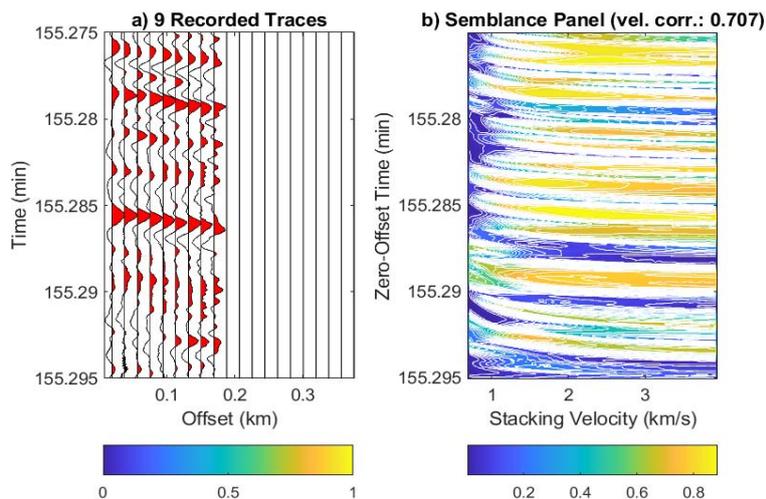

Figure 3: Diving wave events recorded by vertical-component geophones during the Falcon 9 launch on July 11, 2024. Nine geophones were about 7 km from the launch pad: a) recorded seismograms at 9

geophones, b) semblance image from a). The null traces 10-18 in a) will be filled in by extrapolation, and the events between 155.284 and 155.288 minutes will b extrapolated.

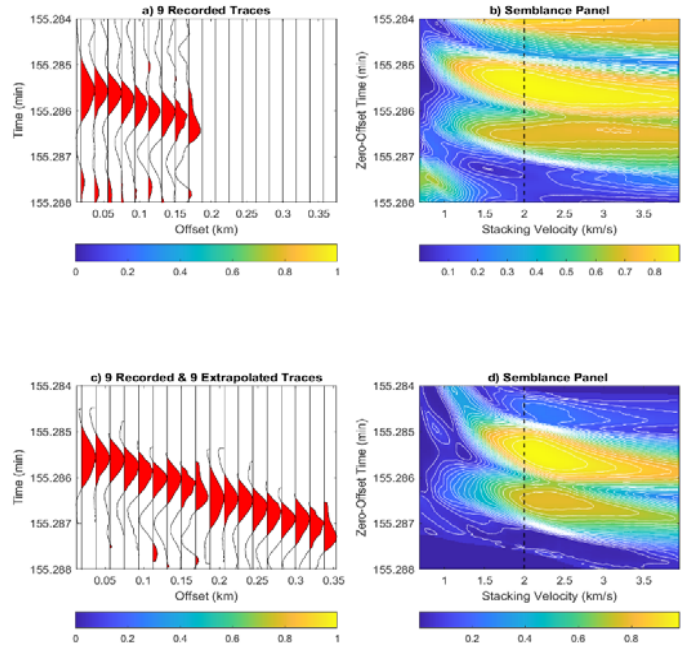

Figure 4: Zoom views of a) windowed events in Figure 3a and b) semblance image in Figure 3b. c) Extrapolated shot gather with 18 traces and d) the corresponding semblance image. The semblance image in d) has significantly less variation of strong semblance values than that in b).

## DISCUSSION AND SUMMARY

The extrapolation procedure assumes a harmonic plane wave propagating across the array where there are no significant lateral changes in velocity. This is not a problem for low-frequency surface waves and short recording arrays. If there are significant lateral velocity variations then the events can be straightened by appropriate time shifts, and the interference of other modes might be eliminated by appropriate 2D filters. Once an initial estimate of the velocity model is made, then the interfering modes can be predicted and further suppressed. This procedure can be performed iteratively until convergence to an acceptable model.

The extrapolation procedure described by equation 3 is a data-driven procedure that only uses the time shift between the nearest and farthest-offset traces for extrapolation. This compares to wavefield extrapolation of reflection data (Yilmaz, 2001) which assumes a velocity model. Appending the extrapolated data to the recorded data and the application of semblance does not reduce the variance of the random noise in the data. However, it can significantly reduce the variance of the semblance contours so that a more reliable estimate of apparent velocity is obtained. This reduction is inversely proportional to the width of the extrapolated recording array. Wavefield extrapolation for migration (Yilmaz, 2001) is downward continuation of recorded reflections in an assumed velocity model, while equation 3 is a sideways extrapolation of recorded data in a model with negligible lateral velocity variations.

If a recording array only has a few traces with erroneous time shifts, then accurate extrapolation of the traces will likely not be possible. The extrapolation procedure can be applied to reflection events where the near-offset data are hyperbolic but the far-offset data have a linear moveout. That is, window the events with a linear moveout and use them to append to the original data. Also, flattening procedures can be used to flatten the reflection events prior to extrapolation. De-flattening these data can then be applied to the extrapolated data.

## APPENDIX

Figure 5 depicts a simple example where extrapolation provides a smaller variance of semblance velocities. In this case, there are 8 traces with pulsed wavelets and a flat moveout. There is noise on traces #2 and #6 which have flipped the polarity of the pulsed amplitudes. The simplified semblance formula (https://wiki.seg.org/wiki/Semblance) is

$$sem(v, t_0) \quad = \quad \frac{1}{N} \frac{(\sum_{i=1}^{N} d(x_i, t_i))^2}{\sum_{i=1}^{8} d(x_i, t_i)^2}, \tag{5}$$

where $t_i$ is the time along the red line in Figure 5 at the *ith* trace with offset $x_i$, and v is the trial moveout velocity which is equal to the slope $1/v$ of the red line. The number of traces is N = 8 and $t_0$ is the zero-offset time where the red line intercepts the #1 trace. Note, the extrapolated traces 5-8 are replicated versions of traces 1-4. In practice, the amplitude values are summed over a fat line with a width of about a half period but we use a summation over a single line to simplify the discussion.

If traces 1-4 are the recorded traces then the semblance value computed from the above formula is

$$sem(v, t_0) \quad = \quad \frac{1}{4} \frac{(1+1)^2}{(1^2+1^2+1^2+1^2)^2} = 1/16, \tag{6}$$

which is the same semblance value 1/16 computed for summation along a line that is horizontal with $1/v = \infty$. In this case, there is a wide range of velocities that give the same semblance value. On the other hand, if traces 1-4 were used to fill in the trace positions 5-8 in Figure 5, then the last trade at position #8 would have a zero value; the semblance value for a horizontal line will be greater than that for the red line in Figure 5. Therefore extrapolated traces will lead to computed semblance velocities with smaller variances compared to those computed from traces 1-4

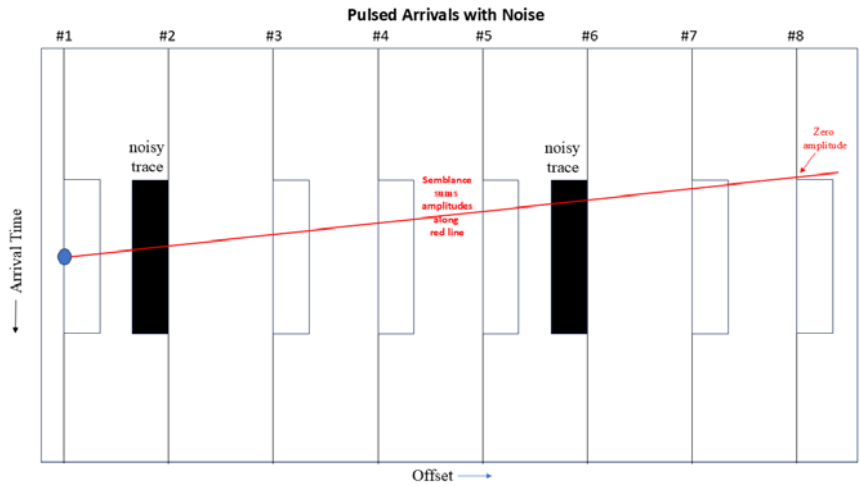

Figure 5: Traces with a flat moveout and the negative amplitudes (solid black) represent noise in the traces. Note, the extrapolated traces 5-8 are duplicated versions of traces 1-4